\documentclass[aps,reprint,prd,nofootinbib,noshowkeys,longbibliography,floatfix,superscriptaddress,preprintnumbers]{revtex4-1}
\usepackage[utf8]{inputenc} % for umlauts etc
\usepackage{hyperref} % for hyperlinks
\usepackage{amsmath} % ams stuff for maths
\usepackage{amsfonts}
\usepackage{amsthm}
\usepackage{amssymb}
\usepackage{graphicx} % for pics
\usepackage{color} % for coloured comments
\usepackage[sort&compress]{natbib}
\usepackage[normalem]{ulem}

\DeclareMathOperator{\tr}{Tr}

\begin{document}

\title{Classical production of 't~Hooft-Polyakov monopoles from magnetic fields}
\date{November 14, 2019}

\author{David L.-J. Ho}
\email{d.ho17@imperial.ac.uk}
\author{Arttu Rajantie}
\email{a.rajantie@imperial.ac.uk}
\affiliation{Department of Physics, Imperial College London, SW7 2AZ, UK}
\preprint{IMPERIAL-TP-2019-DH-02}

\begin{abstract}
We show that in the SU(2) Georgi-Glashow model, ’t Hooft-Polyakov monopoles are produced by a classical instability in magnetic fields above the Ambjørn-Olesen critical field, which coincides approximately with the field at which Schwinger pair production becomes unsuppressed. Below it, monopoles can be produced thermally, and we show that the rate is higher than for pointlike monopoles by calculating the sphaleron energy as a function of the magnetic field. The results can be applied to production of monopoles in heavy-ion collisions or in the early Universe.
\end{abstract}

\maketitle

\section{Introduction}

Magnetic monopoles --- hypothesised particles carrying magnetic charge --- arise as topological solutions to the field equations of certain 
non-Abelian gauge theories \cite{thooft1974magnetic, polyakov1974particle}. It has been proposed \cite{roberts1986dirac, gould2019schwinger, gould2018worldline} that heavy-ion collisions may be the most promising terrestrial method of creating these currently undetected particles, because of the very strong magnetic fields they generate.
In order to gain meaningful information from past \cite{he1997search} and future \cite{acharya2019magnetic} searches, an understanding of monopole production mechanisms is vital.

Production of magnetic monopoles in a strong magnetic field $B_{\rm ext}$ is the electromagnetic dual of production of electrically charged particle-antiparticle pairs in strong electric fields. It occurs because in the presence of a uniform magnetic field, the state with no monopoles is not the true ground state and can therefore decay by producing monopole-antimonopole pairs. In weak fields and at low temperatures, this happens through Schwinger pair creation~\cite{Sauter:1931zz,Schwinger:1951nm}, which means quantum tunneling through the Coulomb potential barrier. It can be described with four-dimensional instanton solutions, and the rate $\Gamma$ of the process is given by the instanton action $S_{\rm inst}=\pi M^2/q_{\rm m}B_{\rm ext}-q_{\rm m}^2/4$, where $M$ is the monopole mass and $q_{\rm m}$ is the magnetic charge of the monopole, through $\Gamma\propto \exp(-S_{\rm inst})$~\cite{affleck1981monopole,affleck1981pair}. At higher temperatures, there is sufficient energy available for the monopoles to cross the potential barrier classically, and then the rate is determined by the energy $E_{\rm sph}$ of the three-dimensional sphaleron configuration~\cite{manton1984saddle} through its Boltzmann weight $\Gamma\propto \exp(-E_{\rm sph}/T)$.

The instantons and sphalerons that describe production of pointlike magnetic monopoles have been previously studied in Refs.~\cite{affleck1981pair, gould2017thermal,gould2018mass,gould2018worldline,gould2019schwinger}. \footnote{Monopole production in a nonabelian gauge theory was also discussed qualitatively in Ref.~\cite{Garriga:1995fv} in the context of Nambu monopoles in electroweak theory.}  In this letter we extend these analyses to solitonic 't~Hooft-Polyakov monopoles, which is important because that is the form in which monopoles appear in many particle physics models and because earlier results~\cite{gould2019schwinger} have shown that the pointlike approximation fails at relativistic collision energies. We show that the sphaleron energies are lower than for pointlike monopoles, and that in sufficiently strong magnetic fields, the potential barrier disappears completely and therefore monopole production takes place through a classical instability and is unsuppressed even at zero temperature.

In fact, this instability corresponds to the well-known Ambj{\o}rn-Olesen instability in strong magnetic fields~\cite{ambjorn1988antiscreening,ambjorn1988electroweak},
which occurs above the critical field strength
\(B_\mathrm{crit} = m_\mathrm{v}^2 / g\), where \(m_\mathrm{v}\) is the mass of the charged vector bosons and \(g\) is the electric charge. Ambj{\o}rn and Olesen found that in the electroweak theory the instability leads to the formation of a lattice of vortex lines, but we demonstrate than in the Georgi-Glashow model it, instead, leads to monopole pair production.

\section{Theory}

We work in the Georgi-Glashow model \cite{georgi1972unified} consisting of an SU(2) gauge field \(A_\mu\) with an adjoint scalar field \(\Phi\): the continuum Lagrangian is
\begin{equation} \label{eq:georgiGlashowAction}
    \mathcal{L} = -\frac{1}{2} \tr (F_{\mu \nu} F^{\mu \nu}) + \tr(D_\mu \Phi D^\mu \Phi) - V(\Phi),
\end{equation}
where
\begin{align}
    D_\mu \Phi^a &:= \partial_\mu \Phi^a + i g \varepsilon^{abc} A_{\mu}^b \Phi^c, \\
    F_{\mu \nu}^a &:= \partial_\mu A_\nu^a - \partial_\nu A_\mu^a + i g \varepsilon^{abc} A_\mu^b A_\nu^c, \\
    V(\Phi) &:= \lambda\left(\tr(\Phi^2) - v^2\right)^2.
\end{align}
The theory has
two dimensionless parameters: the gauge coupling \(g\) and the scalar field self-coupling \(\lambda\), and the scalar field vacuum expectation value (VEV) \(\sqrt{2}v\), which sets the scale.

We focus on static solutions to the field equations, so we are free to work in the 3D theory where all time derivatives, along with the timelike components of the gauge field, vanish. To perform numerical calculations we discretise the action, restricting it to lattice points \(\vec{x} = (n_x, n_y, n_z) a\), where \(n_x, n_y, n_z\) are integers and \(a\) is the lattice spacing. The scalar field \(\Phi(\vec{x})\) is defined on lattice sites, whilst the gauge field is defined via link variables \(U_i(\vec{x})\): in units where \(a = 1\) the discretised energy density is
\begin{widetext}
\begin{equation} \label{eq:latticeLagrangian}
\begin{split}
    \mathcal{E}_\mathrm{lat} = \frac{2}{g^2} \sum_{i < j}[2 - \tr U_{ij}(\vec{x})] + 2 \sum_i \left[\tr \Phi(\vec{x})^2 - \tr \Phi(\vec{x}) U_i(\vec{x}) \Phi(\vec{x} + \hat{\imath}) U^\dagger_i(\vec{x}) \right] + V(\Phi),
\end{split}
\end{equation}
\end{widetext}
using \(U_{ij}\) to denote the standard Wilson plaquette. The sum of this over all lattice sites \(E_\mathrm{lat} = \sum_{\vec{x}} \mathcal{E}_\mathrm{lat}\) is the quantity we extremise. 

The magnetic field corresponding to the residual U(1) symmetry is given through lattice projection operators~\cite{davis2000topological}: defining the projection operator \(\Pi_+ := \tfrac{1}{2}(\mathbf{1} + \Phi / |\Phi|)\), the projected link variable is
\begin{equation}
    u_i(\vec{x}) := \Pi_+(\vec{x}) U_i(\vec{x}) \Pi_+(\vec{x} + \hat{\imath}).
\end{equation}
This gives an abelian field strength tensor
\begin{equation}
    \alpha_{ij} := \frac{2}{g} \arg \tr u_i(\vec{x}) u_j(\vec{x} + \hat{\imath}) u_i^\dagger(\vec{x} + \hat{\jmath}) u_j^\dagger(\vec{x}),
\end{equation}
from which the magnetic field strength may be obtained: \(B_i := \epsilon_{ijk} \alpha_{jk}/2a^2\).

It is defined modulo $B_{\rm max}\equiv 2\pi/ga^2$, and therefore the magnetic charge, defined by the magnetic field's divergence, is quantised in units of \(4 \pi / g\).

The theory displays spontaneous symmetry breaking, generating a scalar boson mass \(m_\mathrm{s} = 2 \sqrt{\lambda} v\) and charged vector boson masses \(m_\mathrm{v} = \sqrt{2} g v\); there remains an unbroken U(1) symmetry giving a massless photon. 
It admits 't~Hooft-Polyakov monopole solutions \cite{thooft1974magnetic, polyakov1974particle} of magnetic charge \(q_\mathrm{m} = 4 \pi n / g\) for \(n \in \mathbb{Z}\). The classical monopole mass is
\begin{equation} \label{eq:monopoleMass}
    M = \frac{4 \pi m_\mathrm{v}}{g^2} f(\beta),
\end{equation}
where \(\beta := m_\mathrm{s} / m_\mathrm{v} = (2\lambda/g^2)^{1/2}\), and the function \(f(\beta) \sim 1\) for all \(\beta\) \cite{forgacs2005numerical}. The monopole has a characteristic radius \(r_\mathrm{m} \sim m_\mathrm{v}^{-1}\).

We look for a sphaleron configuration --- an unstable static solution with a single negative mode --- in the presence of a non-zero magnetic field $B_{\rm ext}$, which corresponds to the top of the barrier between the uniform field state and the state with a monopole-antimonopole pair.\footnote{Note that the theory also has the Taubes sphaleron solution~\cite{taubes1982existence}, which is physically different and which we are therefore not investigating in this work.} The energy $E_{\rm sph}$ of this configuration gives the minimum energy that is needed to move from the uniform field state to a state with a monopole-antimonopole pair.

If the external field \(B_\mathrm{ext}\) is weak enough for the sphaleron size to be large compared to the monopole size, a pointlike monopole approximation is valid~\cite{gould2017thermal,gould2018worldline}. Accounting for both the Coulomb force and short-range interactions due to the charged vector bosons and scalar particles, the monopole-antimonopole potential (for poles aligned in isospace) can be estimated, in terms of pole separation \(r\), as~\cite{saurabh2017monopole}
\begin{equation} \label{eq:pointlikePotential}
    V_\mathrm{m\bar{m}} = -\frac{1}{4 \pi r} \left[1 + 2 \mathrm{e}^{-m_\mathrm{v} r} + \mathrm{e}^{-m_\mathrm{s} r}(1 - \mathrm{e}^{-m_\mathrm{v} r} ) \right].
\end{equation}
The scalar and massive vector bosons result in a short-range attractive force between an untwisted pair, lowering the interaction energy compared to the Coulomb case. Eq.~\eqref{eq:pointlikePotential} is valid providing the monopole separation is large compared to the core size: \(r \gg r_\mathrm{m}\).
In this case the sphaleron energy $E_\mathrm{sph}$ is
the maximum of the function
\begin{equation} \label{eq:pointlikeSphaleronEnergy}
    E(r) = 2M - q_\mathrm{m} B_\mathrm{ext} r + V_\mathrm{m \bar{m}}(r),
\end{equation}
and we denote by $r_\mathrm{sph}$ the position of this maximum,
$E_\mathrm{sph}=E(r_\mathrm{sph})$.
The pointlike approximation provides a useful comparison to our results; it is expected to break down when \(r_\mathrm{sph} \approx r_\mathrm{m}\).

\section{Numerical Methods}

We search for saddle points of the discretised energy functional \(E_\mathrm{lat} = \sum_{\vec{x}} \mathcal{E}_\mathrm{lat}\). This is a considerably harder computational task than finding a minimum-energy solution. We achieve this using a modified gradient flow algorithm  proposed by Chigusa \emph{et al.}~\cite{chigusa2019bounce}. This converges on saddle points by including a term in each flow iteration that lifts the negative mode, and is summarised briefly below; for full details consult the referenced paper.

If we denote the set of all field and link variables by \(X\), and an individual field or link variable by \(X_\alpha\), a na\"{i}ve gradient flow update is

\begin{equation}
    X_\alpha(\tau + \delta \tau) = X_\alpha(\tau) - \partial_\alpha E_{\rm lat} \delta \tau,
\end{equation}
where \(\tau\) denotes `flow time' and
\begin{equation} \label{eq:naiveGradientFlow}
    \partial_\alpha E_{\rm lat} := \frac{\partial E_{\rm lat}}{\partial X_\alpha}.
\end{equation}
Instead of this, Chigusa \emph{et al.}'s algorithm~\cite{chigusa2019bounce} uses the update
\begin{equation} \label{eq:chigusaFlow}
    X_\alpha(\tau + \delta \tau) = X_\alpha(\tau) - \left[\partial_\alpha E_\mathrm{lat}
    - k G_\alpha \sum_\beta (\partial_\beta E_{\rm lat})G_\beta \right] \delta \tau,
\end{equation}
where \(k\) is a real constant, and \(G\) is a field on the lattice, normalised such that \(\sum_\beta G_\beta G_\beta = 1\). In Ref.~\cite{chigusa2019bounce}, it is shown that any fixed point of the flow \eqref{eq:chigusaFlow} is also a stationary point of \(E_\mathrm{lat}\), and that for suitably chosen \(k\) and \(G\), this stationary point will be a saddle point.

The ideal choice for \(G\) would be directly proportional to the negative mode of the saddle point field configuration --- then the modified gradient flow would descend along all positive modes and ascend along the negative mode to the sphaleron. However, the sphaleron and its negative mode are obviously unknown. We thus use a heuristic prescription to choose \(G\) as close as possible to the true negative mode: starting from an initial configuration sufficiently close to the sphaleron, standard gradient flow using Eq.~\eqref{eq:naiveGradientFlow} will minimise along the positive modes whilst continuing to flow the system along the negative mode. The point along the flow closest to the saddle point may be identified by considering the sum of the squares of the gradients at all lattice points. When this point is reached, the deviation from the true saddle point will be largely along the negative mode, and thus the modified flow \eqref{eq:chigusaFlow} is likely to converge along the saddle point. The normalised gradient at this point is used as \(G\) in our calculations:

\begin{equation}
G_\alpha=\frac{\partial_\alpha E_{\rm lat}}{\left(\sum_\beta \partial_\beta E_\mathrm{lat} \partial_\beta E_\mathrm{lat}\right)^{1/2}}
\end{equation}

evaluated at the point along the standard gradient flow trajectory closest to the saddle point. Using the technique described above we have been able to find the solitonic equivalent of the sphaleron studied in Refs.~\cite{gould2017thermal, gould2018worldline} on the lattice.

Our calculations were carried out on a periodic \(64^3\) lattice using the \textsc{LAT}field2 C++ library \cite{daverio2015latfield} for parallelisation.
A Barzilai-Borwein adaptive step size \cite{barzilai1988step} 
was used to speed convergence. 
A non-zero magnetic field was introduced by initial conditions with total magnetic flux $48\pi/g$, giving a uniform magnetic field strength $B_{\rm ext}\approx 0.037/ga^2$.
As the periodic boundary conditions quantise the flux through the lattice in units of \(4 \pi/g\), an iterative gradient flow evolution is unable to change the flux through the boundary (without moving monopoles to the edges). This has the effect of fixing the asymptotic magnetic field at the desired value.
The magnetic field in units of $m_\mathrm{v}^2$ was varied incrementally by changing the scalar VEV $\sqrt{2}v$, keeping $\beta$ constant. 
Three values of \(\beta\) were investigated: \(\beta = 0.5\), \(\beta = 1\) and \(\beta = 2\).

\section{Sphaleron for 't~Hooft-Polyakov monopoles}

\begin{figure}
    \centering
    \includegraphics[width=0.5\textwidth]{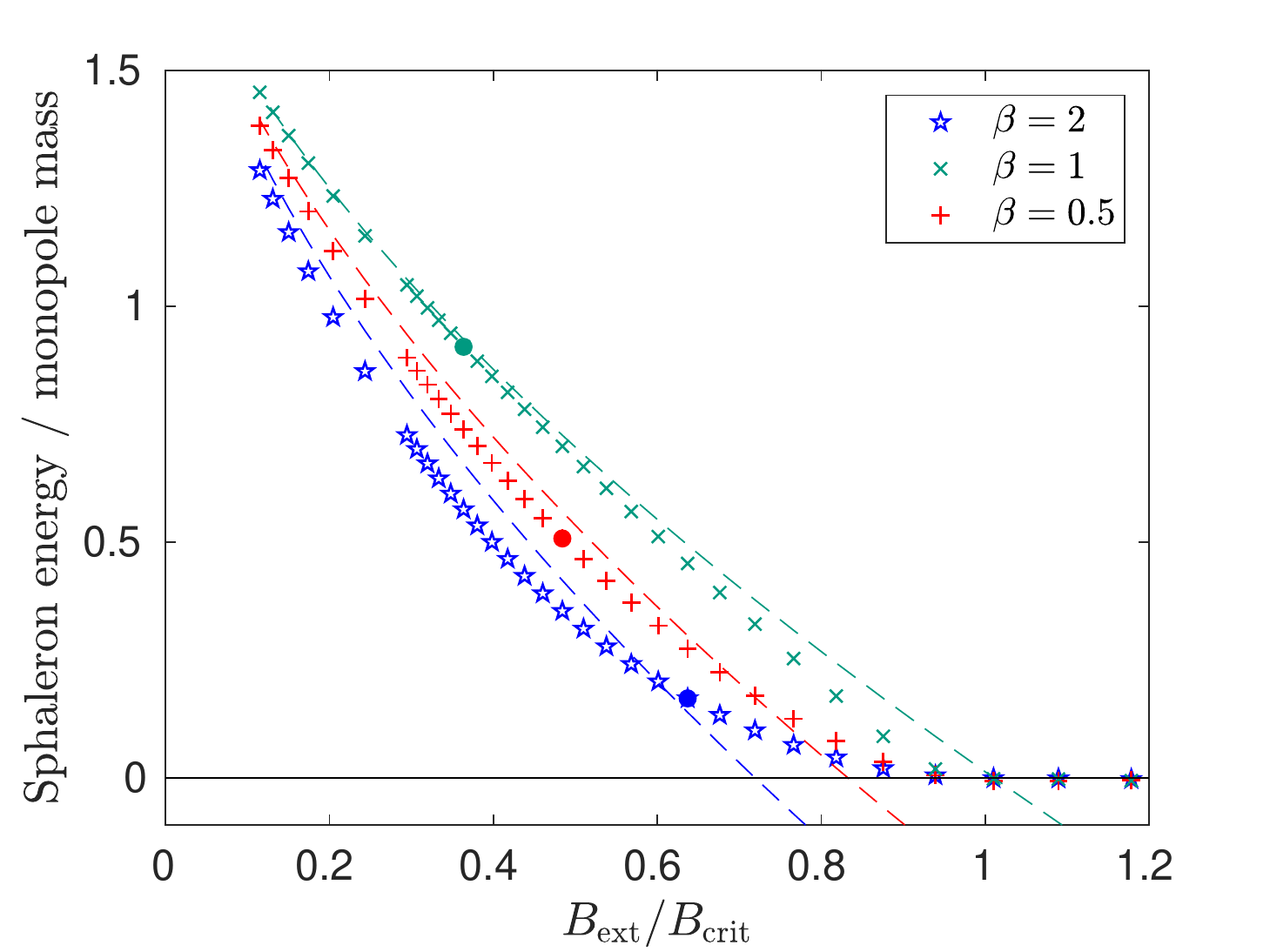}
    \caption{Plot of sphaleron energy against field strength for different values of \(\beta = m_\mathrm{s} / m_\mathrm{v}\). The  dashed lines indicate the predicted sphaleron energy assuming pointlike monopoles (Eq.~\eqref{eq:pointlikeSphaleronEnergy}). The solid circles indicate the field strength at which the sphaleron ceases to contain separated magnetic charges.}
    \label{fig:fieldEnergyPlot}
\end{figure}

\begin{figure*}
    \centering
    \begin{tabular}{c@{}c@{}c@{}}
        \includegraphics[width=0.33\textwidth]{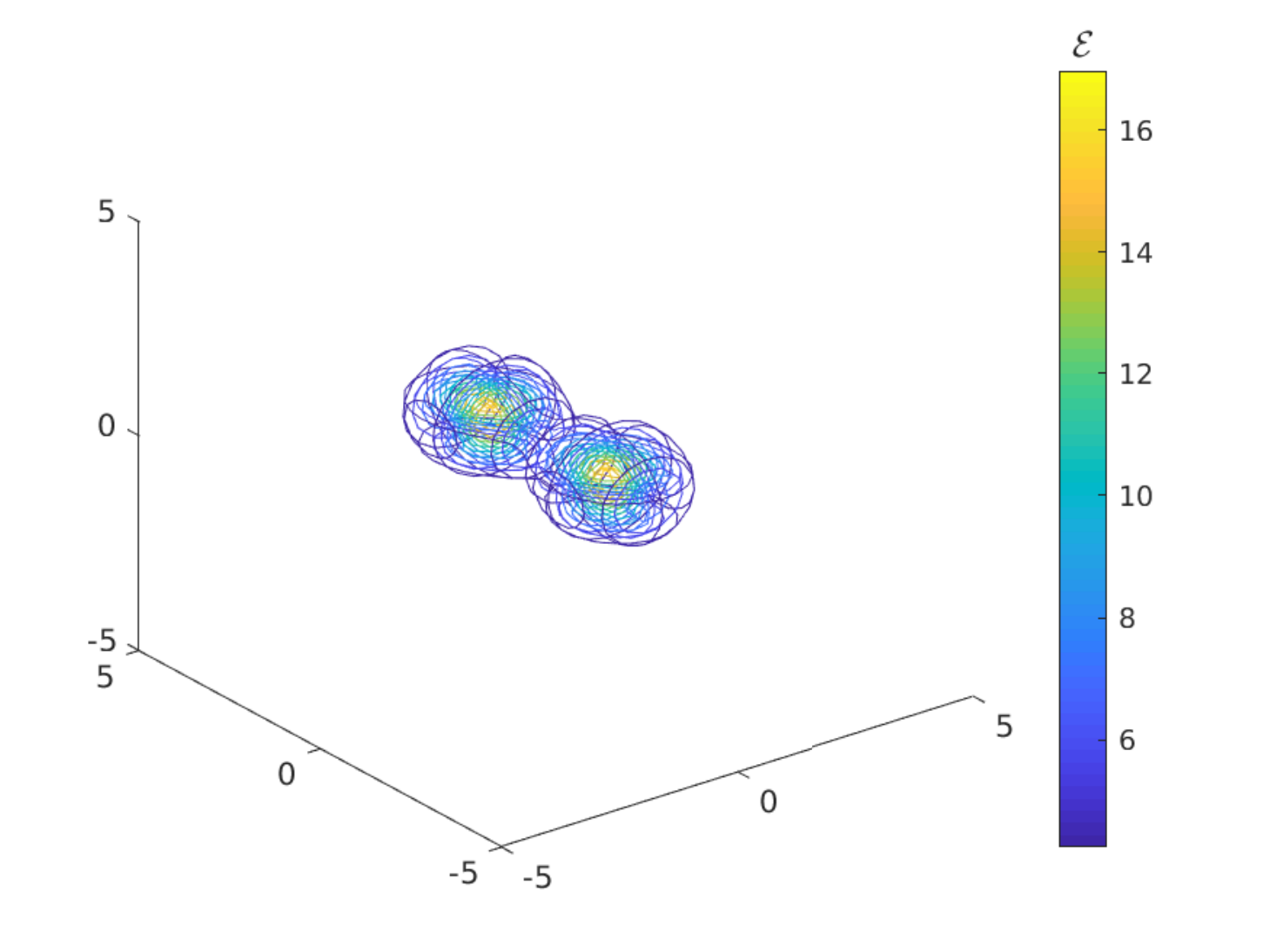} &
        \includegraphics[width=0.33\textwidth]{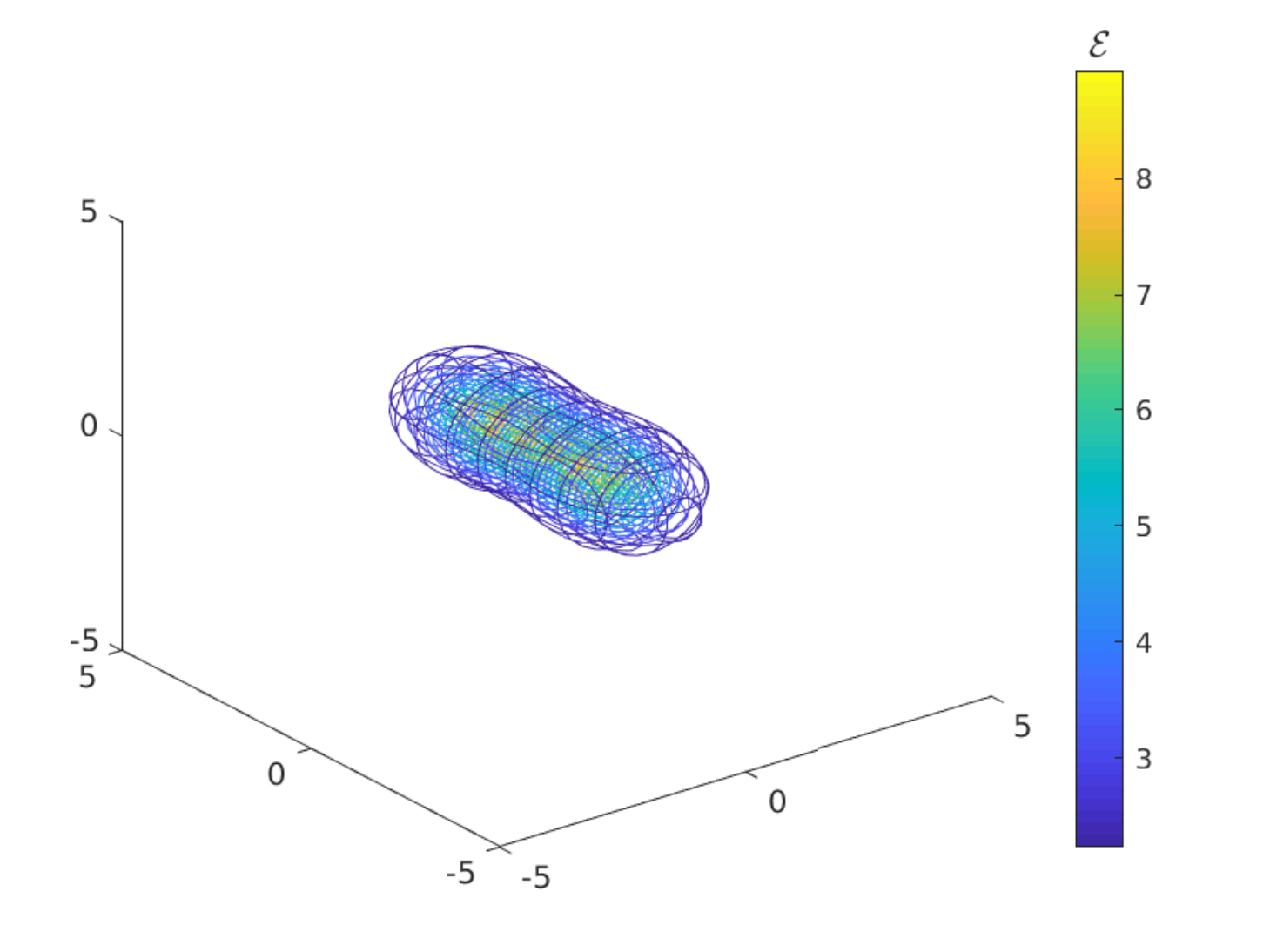} &
        \includegraphics[width=0.33\textwidth]{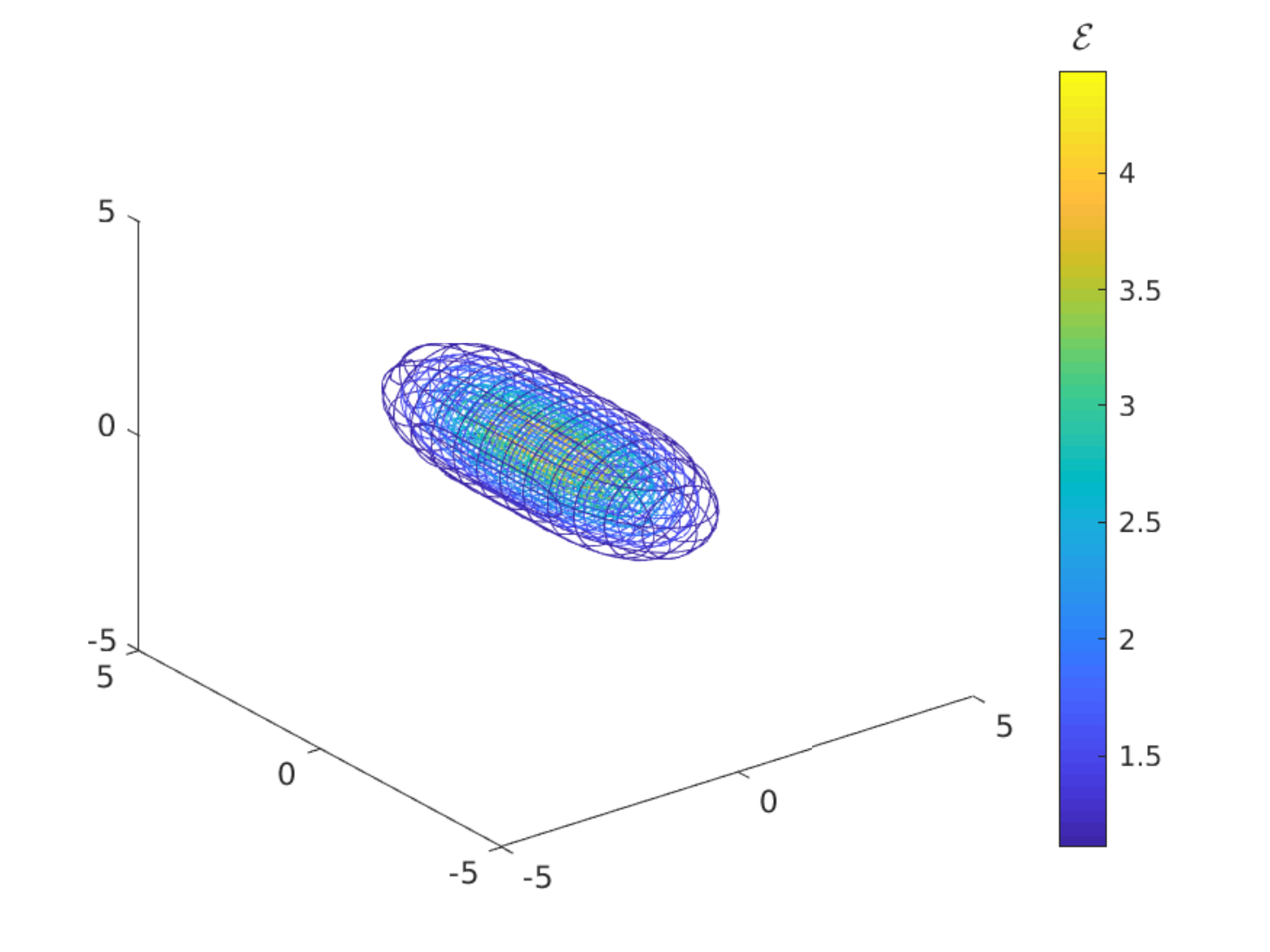} \\
        \includegraphics[width=0.33\textwidth]{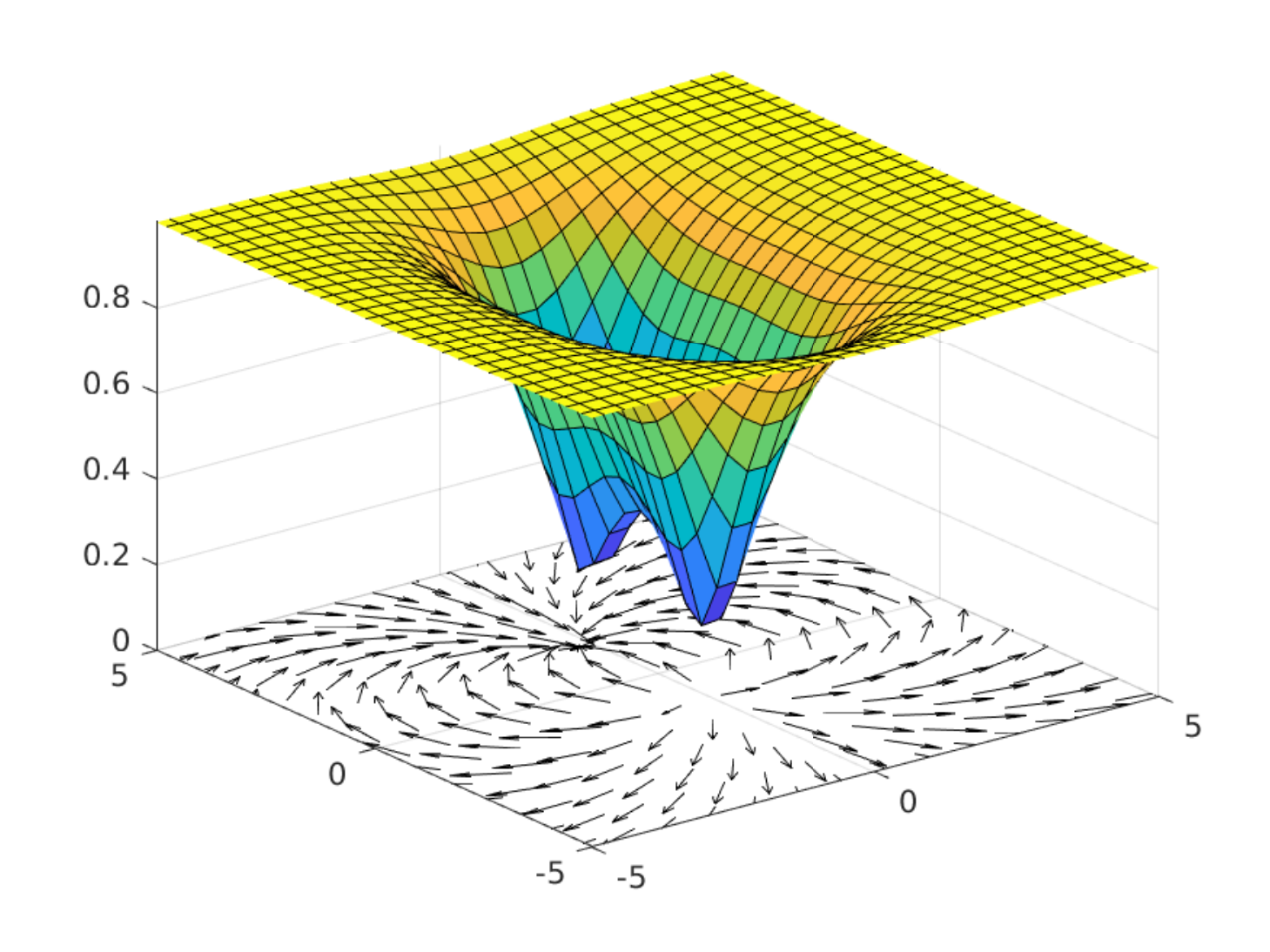} &
        \includegraphics[width=0.33\textwidth]{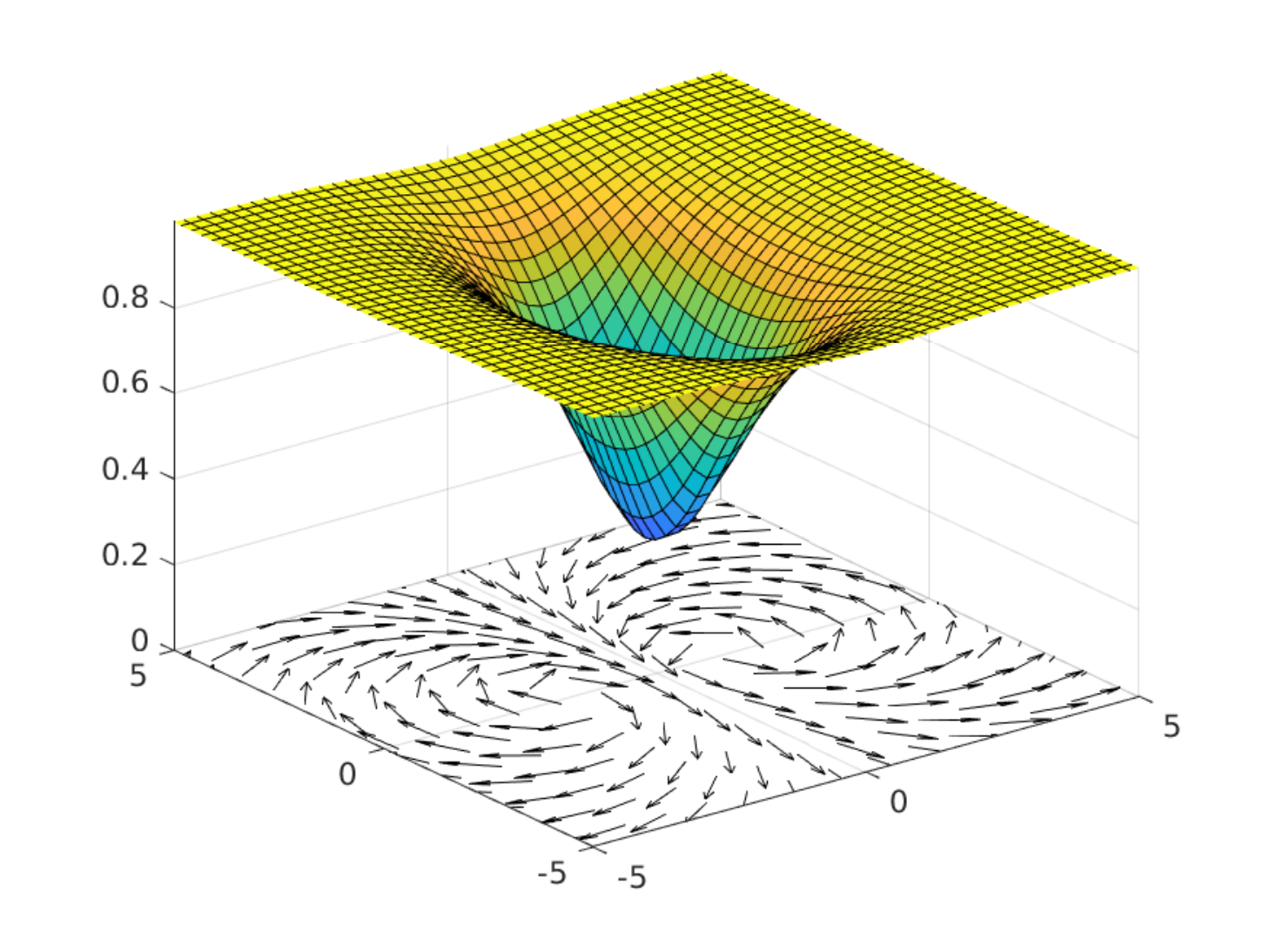} &
        \includegraphics[width=0.33\textwidth]{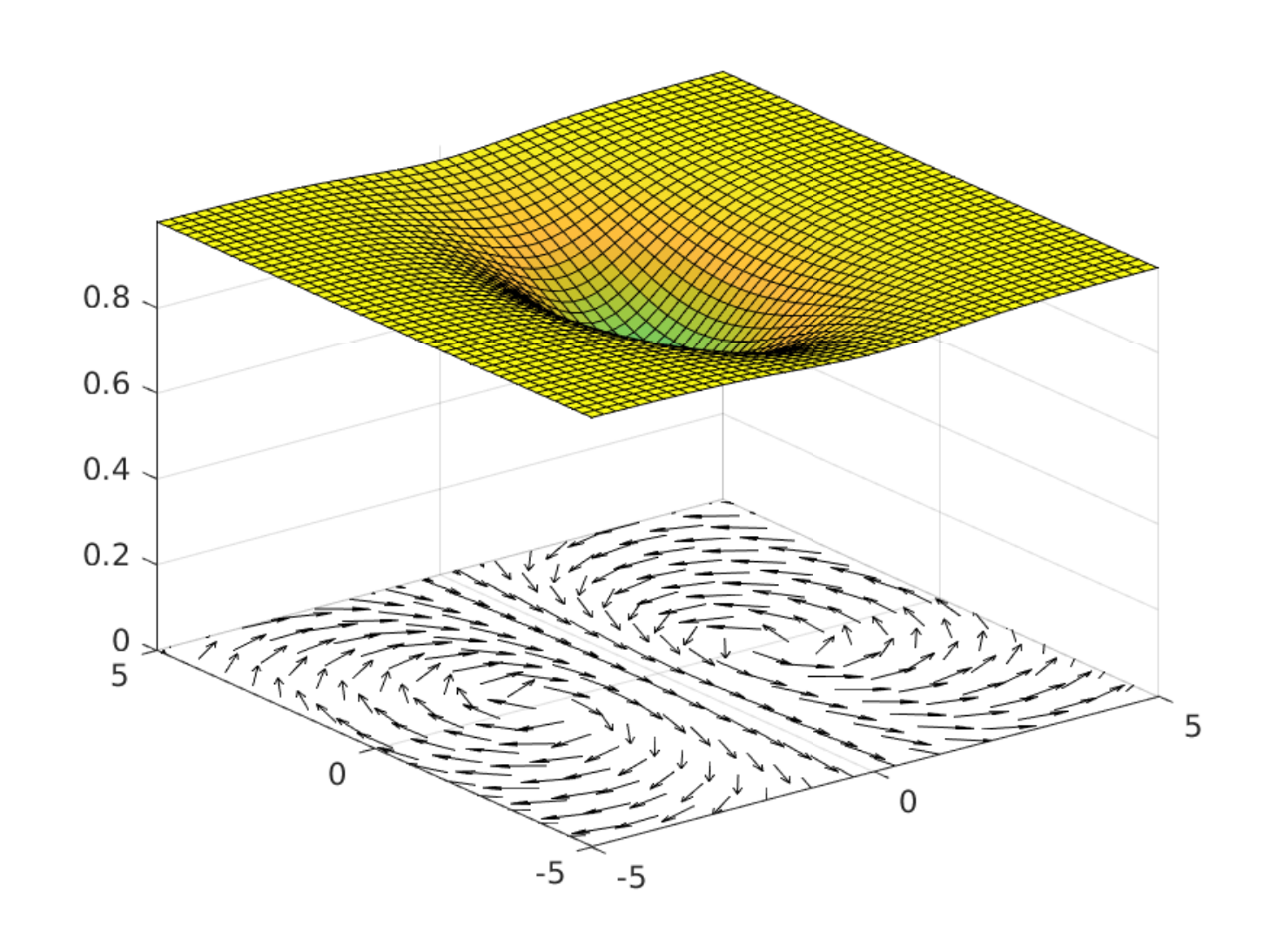} \\
        {(}a{)} \(B/B_\mathrm{crit} = 0.29\) &
        {(}b{) \(B/B_\mathrm{crit} = 0.60\)} &
        {(}c{) \(B/B_\mathrm{crit} = 0.82\)}
    \end{tabular}
    \caption{Visualisations of sphaleron solutions for subannihilating (a) and superannihilating (b, c) magnetic field strengths. Top plots show energy density contours in units of \(m_\mathrm{v}^4\) in 3D space. Bottom plots show slices in a plane parallel to the magnetic field intersecting the sphaleron at its centre: the surface gives the scalar field magnitude in units of its VEV, whilst the vector plots give the direction of the magnetic field (with the background subtracted) through the same slice. All plots shown have \(m_\mathrm{v} = m_\mathrm{s}\). Spatial axes have units of \(m_\mathrm{v}^{-1}\).}
    \label{fig:sphaleronPlots}
\end{figure*}

For weak magnetic fields the sphaleron bears a clear resemblance to the pointlike approximation of a monopole and an antimonopole separated along the direction of the external field. The magnetic charge is non-zero in two cubes lying on a line parallel to the field axis, and the magnitude of the scalar field has two minima, at the same points (due to discretisation effects the scalar field does not vanish). An example of a sphaleron solution with separated magnetic charges is shown in Fig.~\ref{fig:sphaleronPlots}(a).

Fig.~\ref{fig:fieldEnergyPlot} shows the dependence of sphaleron energy on external field strength. As the field strength increases, the energy barrier to monopole production lowers. For fields well below the critical field strength this fits well to the sphaleron energy for pointlike charges \eqref{eq:pointlikeSphaleronEnergy}. As the field increases the calculated sphaleron energy dips below the point particle prediction. This is likely due to the effects of partial cancellation between the overlapping monopole cores (as observed in Ref.~\cite{saurabh2017monopole}).

As the field increases further, 
the distance between
the positive and negative magnetic charges decreases and eventually they cancel each other. We refer to this phenomenon as `annihilation', though it is not a dynamical process.
The higher the value of \(\beta\), the stronger the field required to annihilate the monopoles. (see Fig.~\ref{fig:fieldEnergyPlot}). There is no visible discontinuity in energy at annihilation.

Above the annihilation threshold, the scalar field has only one minimum, and the magnetic charge is zero everywhere.
The sphaleron still has a non-zero magnetic dipole moment, originating
 from an axisymmetric ring of electric current density centred about the minimum of the scalar field (see Fig.~\ref{fig:sphaleronPlots}(b,c)). For fields slightly stronger than the annihilation field strength the energy density contours of the sphaleron continue to define a peanut-like shape with two separate maxima (Fig.~\ref{fig:sphaleronPlots}(b)). For very high magnetic fields energy contours are pill-shaped (Fig.~\ref{fig:sphaleronPlots}(c)).

The sphaleron energy decreases monotonically with increasing \(B_\mathrm{ext}\) until it reaches zero, where it plateaus (see Fig.~\ref{fig:fieldEnergyPlot}). At this point the saddle point configuration transitions to the vacuum configuration with only the homogeneous background magnetic field present. 
From Fig.~\ref{fig:fieldEnergyPlot} it can be seen that the field strength where this happened is independent of \(\beta=m_\mathrm{s} / m_\mathrm{v}\),
and appears to coincide with the Ambj{\o}rn-Olesen critical field \cite{ambjorn1988antiscreening},
\(B_\mathrm{ext} = B_\mathrm{crit} = m_\mathrm{v}^2/g\).

Above \(B_\mathrm{crit}\) there is no energy barrier to the creation of monopole-antimonopole pairs, which suggests that monopole-antimonopole pairs are produced by a classical instability.

\begin{figure*}
    \centering
    \begin{tabular}{c@{}c@{}}
        \includegraphics[width=0.45\textwidth]{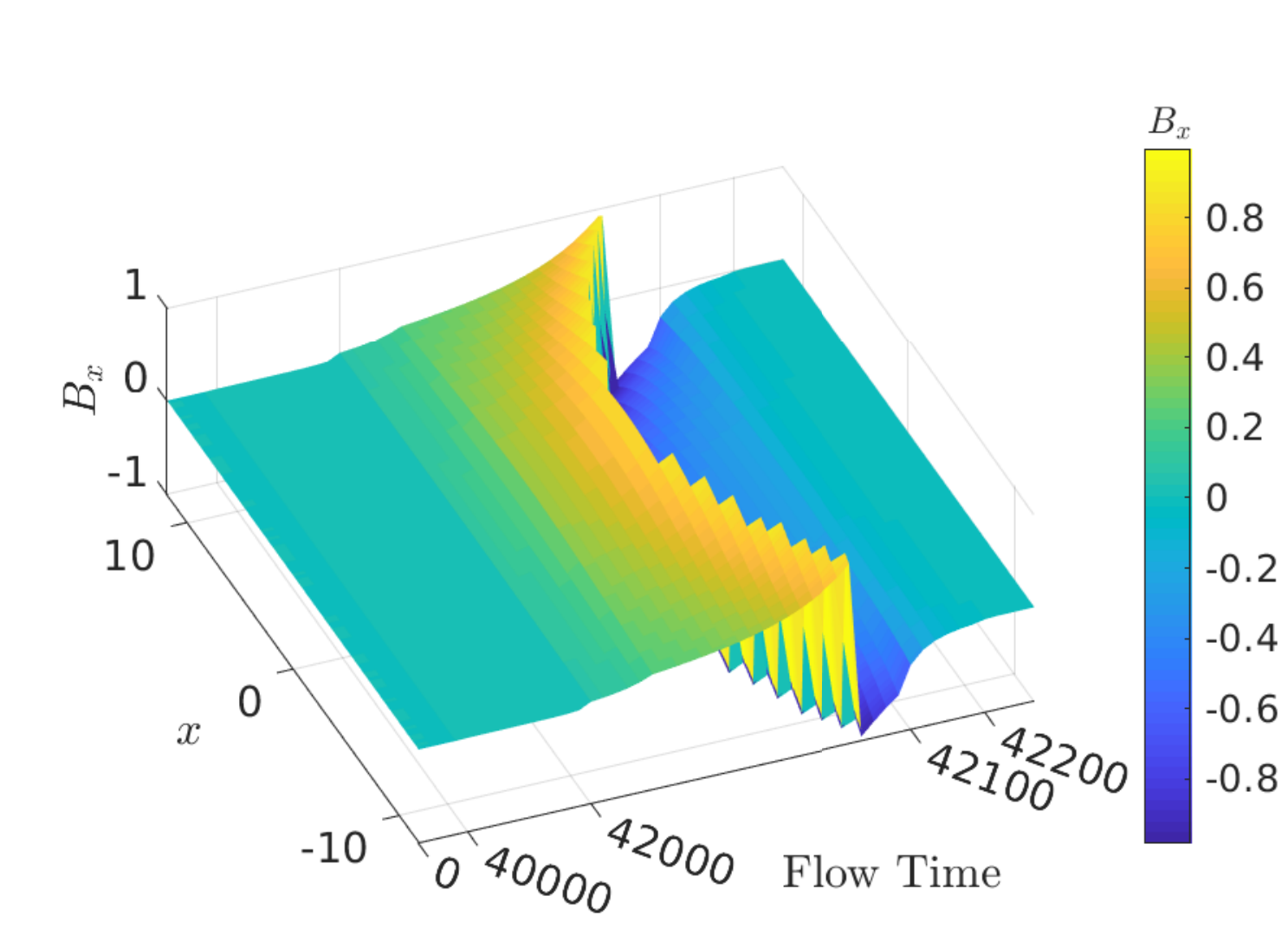} & \qquad
        \includegraphics[width=0.45\textwidth]{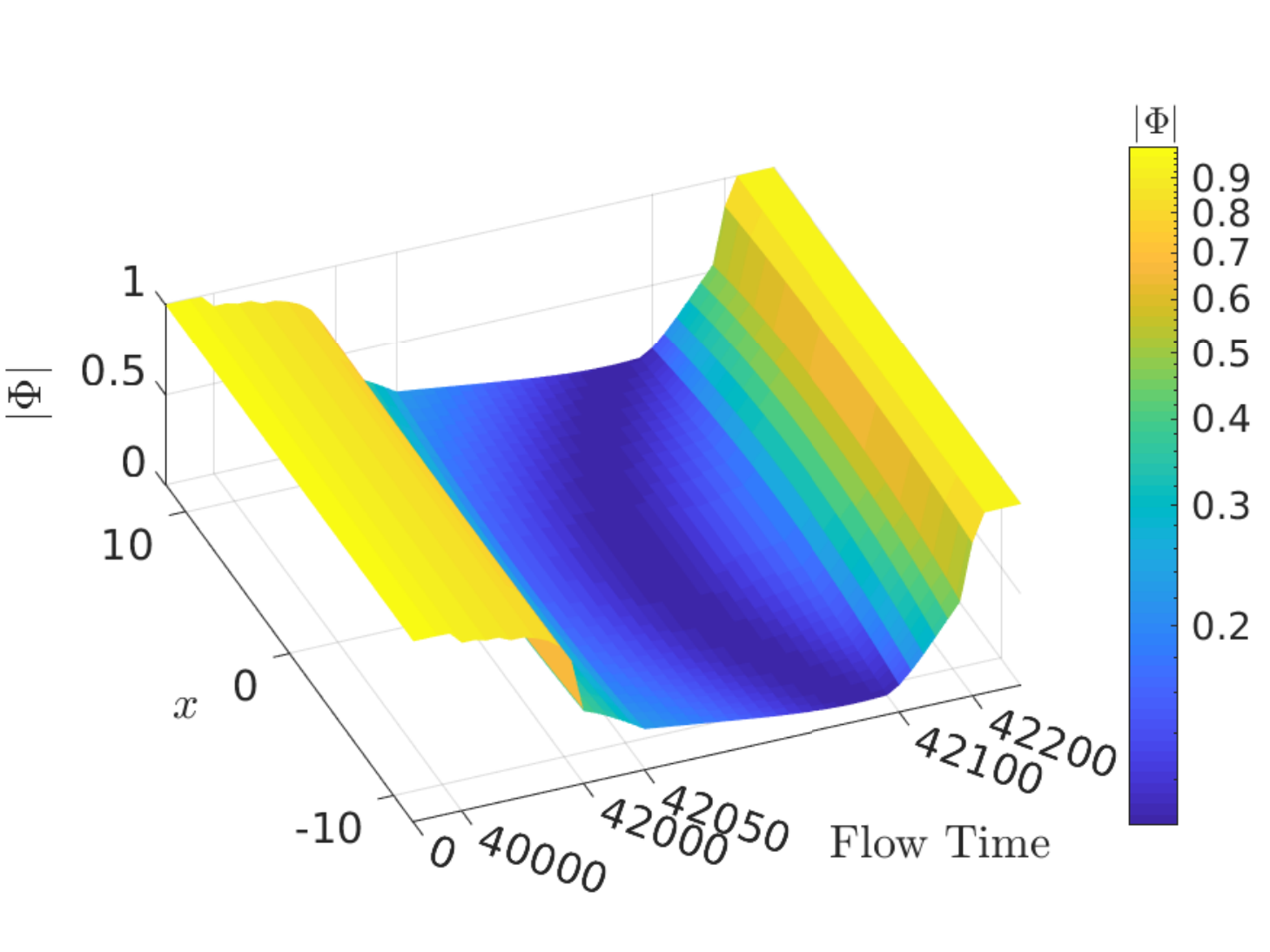}
    \end{tabular}
    \caption{Standard gradient flow evolution of the magnetic field component in the direction of the external field $B_x$ (left) and the scalar field magnitude $|\Phi|$ (right), starting from a supercritical homogeneous magnetic field with a stochastic perturbation and leading to the production of a monopole-antimonopole pair at flow time $42050$.
    Values are taken along a line parallel to the field axis, passing through the cores of the produced monopoles.  The magnetic field is given in units of $B_{\rm max}$, %\(2 \pi / g a^2\), 
    the scalar field magnitude in units of its VEV, and spatial distance in units of \(m_\mathrm{v}^{-1}\). The colours indicate the magnitude of the quantity plotted as shown in the colour bar. Note the uneven scale on the flow time axis.
    The plot shows the magnetic field strength $B_x$ increasing exponentially due to the instability until it reaches $B_{\rm max}$, %\(2 \pi / g a^2\), 
    at which time the monopoles (seen as discontinuities in the magnetic field) form. Once formed, the monopoles rapidly move to the boundary where they annihilate, allowing the system to settle in the new equilibrium state with lower magnetic field.
    }
    \label{fig:flowEvolutionPlots}
\end{figure*}

We investigated this hypothesis by performing standard gradient flow from a uniform supercritical background field with small random white noise perturbations in the gauge fields. The results are summarised in Fig.~\ref{fig:flowEvolutionPlots}. A clear instability is seen, as predicted in Ref.~\cite{ambjorn1988antiscreening}, but rather than stabilising to the Ambj{\o}rn-Olesen vortex lattice solution presented in Ref.~\cite{ambjorn1988antiscreening}, the magnetic field continues to become increasingly more localised. The local field strength at the centre of the vortex grows exponentially  until it reaches $B_{\rm max}$, %$2\pi/ga^2$, 
which is the maximum value allowed on the lattice. At this time (\(\approx 42050\) flow time units in Fig.~\ref{fig:flowEvolutionPlots}), a monopole-antimonopole pair is produced. This may be interpreted as the breaking of the vortex.

At the time of the pair production, both the magnetic field and the energy density are highly localised in a vortex line aligned with the external field. Inside the vortex line, the scalar field vanishes (see Fig.~\ref{fig:flowEvolutionPlots}), which restores the SU(2) symmetry locally, and therefore the energy density remains finite $\sim V(0)=\lambda v^4$, in spite of the local magnetic field reaching nominally cutoff-scale values.
When the local magnetic field crosses $B_{\rm max}$, % $2\pi/ga^2$, 
it flips sign, and the vortex line breaks forming monopoles, which quickly move to the edges of the lattice and annihilate, lowering the magnetic flux by $4\pi/g$.

It is interesting to note that the critical field strength we have found for classical monopole pair production agrees almost exactly with the field strength at which quantum Schwinger pair production of pointlike monopoles becomes unsuppressed~\cite{affleck1981monopole,affleck1981pair},
\begin{equation}
    B_{\rm Schwinger}\approx \frac{4\pi M^2}{q_\mathrm{m}^3}
    =f(\beta)^2 B_{\rm crit},
\end{equation} 
where we have used Eq.~(\ref{eq:monopoleMass}).
Though this may seem like an unlikely coincidence, it is not entirely unexpected
because this is the natural field strength given by dimensional analysis.
It suggests that the Schwinger process turns continuously to the classical instability when the field exceeds the critical value.

For comparison, we also carried out a similar calculation in the electroweak theory, where we do find the stable vortex lattice predicted in Ref.~\cite{ambjorn1988electroweak}.
In future work it may be interesting to consider 
modifications of the electroweak theory that contain monopole solutions~\cite{cho1996monopole, ellis2016price}. If a similar phenomenon to ours occurs in these theories at obtainable magnetic fields, future heavy-ion collision data could be used to further constrain these models.

It is also worth considering if the classical production of monopoles could ever be observed in a laboratory. Electroweak theory does not permit solitonic monopoles, so the relevant mass
$m_\mathrm{v}$ is not the electroweak W boson mass, but the mass of the charged gauge bosons associated with the 't~Hooft-Polyakov monopole. Experimental searches for heavy charged bosons give a lower-bound mass of \(5200 \, \mathrm{GeV}\) \cite{tanabashi2018review}, which implies a lower bound on the magnetic field strength required to produce monopoles of \(9 \times 10^{7} \, \mathrm{GeV}^2 \approx 4.5 \times 10^{23} \, \mathrm{T}\). As the fields in current LHC heavy-ion collisions are of order \(1 \, \mathrm{GeV}^2\) \cite{deng2012event}, classical monopole production is impossible to achieve with current technology. However, sufficiently strong magnetic fields may have been present in the early Universe~\cite{Turner:1987bw}.

Furthermore, even if such field strengths could be reached in experiments, one would expect monopoles to be produced by 
Schwinger pair creation at lower field strengths. 
Our results show that at those fields, the thermal Schwinger production rate is  higher than for pointlike monopoles, because the sphaleron energy is somewhat lower (see Fig.~\ref{fig:fieldEnergyPlot}). 
The lower-bound mass for solitonic monopoles is therefore stronger than the \((2 + 2.6 n_\mathrm{D}^{3/2}) \, \mathrm{GeV}\) value (\(n_\mathrm{D}\) denotes the number of Dirac charge quanta) obtained in Ref.~\cite{gould2018mass}.
A key question is whether this enhancement of production rate still holds in spacetime dependent cases such as that investigated in Ref.~\cite{gould2019schwinger}, thought to be valid for ultrarelativistic heavy-ion collisions.

\section*{Acknowledgements}
The authors wish to acknowledge the Imperial College Research Computing Service for computational resources. D.L.-J.H. was supported by a U.K. Science and Technology Facilities Council studentship. A.R. was supported by the U.K. Science and Technology Facilities Council grant ST/P000762/1 and Institute for Particle Physics Phenomenology Associateship.

\bibliography{refs}

\end{document}